\documentclass[aps,prl,twocolumn,superscriptaddress,groupaddress] {revtex4-1} 
\usepackage{amssymb,amsmath,graphicx}
\usepackage{hyperref}

\begin{document}


\title{The dynamical strength of social ties in information spreading}


\author{Giovanna Miritello}
\affiliation{Departamento de Matem\'aticas \& GISC, Universidad Carlos III de Madrid, 28911 Legan\'es, Spain}
\affiliation{Telef\'onica Research, Madrid, Spain}

\author{Esteban Moro}
\affiliation{Departamento de Matem\'aticas \& GISC, Universidad Carlos III de Madrid, 28911 Legan\'es, Spain}
\affiliation{Instituto de Ciencias Matem\'aticas {CSIC-UAM-UCM-UC3M}}
\affiliation{Instituto de Ingenier\'{\i}a del Conocimiento, Universidad Aut\'onoma de Madrid, 28049 Madrid, Spain}

\author{Rub\'en Lara}
\affiliation{Telef\'onica Research, Madrid, Spain}

\date{\today}

\begin{abstract}
We investigate the temporal patterns of human communication and its influence on the spreading of information in social networks. The analysis of mobile phone calls of 20 million people in one country shows that human communication is bursty and happens in group conversations. These features have opposite effects in information reach: while bursts hinder propagation at large scales, conversations favor local rapid cascades. To explain these phenomena we define the dynamical strength of social ties, a quantity that encompasses both the topological and temporal patterns of human communication. \end{abstract}


\maketitle

Quantitative understanding of human communication patterns is of paramount importance to explain the dynamics of many social, technological and economic phenomena \cite{lazer,barrat,castellano,newmansiam}. Most studies have focused on the study of the complex {\em topological patterns} of the underlying contact network (whom we talk to) and its influence in the properties of spreading phenomena in social networks such diffusion of information, innovations, computer viruses, opinions, etc.\ \cite{barrat}. Paradoxically, most of these studies of dynamical phenomena on social networks neglect the {\em temporal patterns} of human communications: humans act in bursts or cascades of events \cite{barabasinature,vazquez,rybski,isella}, most ties are not persistent \cite{kossinets,hidalgo} and communications happen mostly in the form of group conversations \cite{isella,eckmann,qiankun,wu}. However, since information transmission and human communication are concurrent, the temporal structure of communication must influence the properties of information spreading. Indeed, recent experiments of electronic recommendation forwarding \cite{esteban} and simulations of epidemic models on email and mobile databases \cite{vazquez,karsai} found that the asymptotic speed of information spreading is controlled by the bursty nature of human communications that leads to a slowing down of the diffusion. However, although the asymptotic speed is an important property of the propagation of information in social networks, there is still no general understanding of how and what temporal properties of human communication do influence spreading processes and in turn, how they affect the very definition of social interaction.

The answer to this question can be framed in the more general problem of how to model dynamical social networks \cite{kossinets,gautreau}. In most studies, real temporal activity is aggregated over time giving a static snapshot of the social interaction where ties are described by static strengths which do not include information about the temporal aspects of how humans interact. Temporal and topological aspects are therefore disentangled in the analysis. In this letter we merge both aspects in the case of information diffusion by adopting a functional definition of the social ties using the well-known map between dynamical epidemic models and static percolation \cite{newman}. The network is still described by a static graph, but the interaction strength between individuals now incorporates the causal and temporal patterns of their communications and not only on the intensity \cite{onnela}.

To this end we study the mobile communication patterns from a European operator in a single country over a period of 11 months. The data consists of $2\times 10^7$ phone numbers and $7\times 10^8$ communication ties for a total of $9$ billion calls between users.
Call Detail Record (CDR) contains the hashed  number of the caller and the receiver, the time when the call was initiated and the duration of the call. We consider only events in which the caller and the callee belong to the operator under consideration, because of the partial access to the records of other operators. Our data for the connectivity of the social network, the duration of the calls, etc.\ are very similar to the ones reported in previous studies \cite{onnela}.  

\begin{figure}
\includegraphics[width=0.40\textwidth]{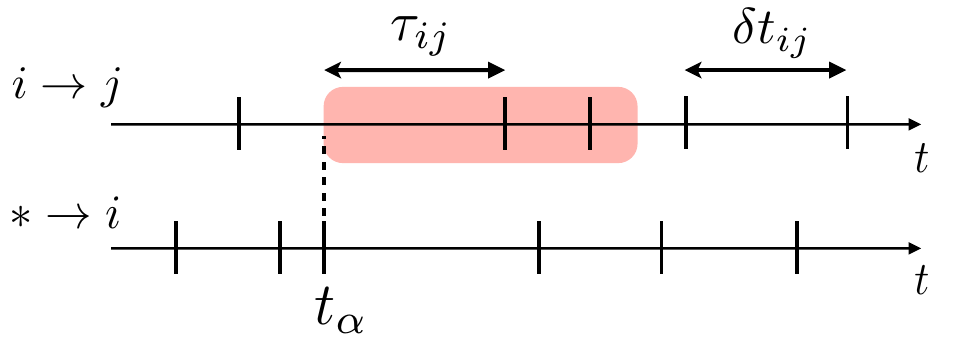}
\caption{(color online) Schematic view of communications events around individual $i$: each horizontal segment indicates an event between $i\to j$ (top) and $*\to i$ (bottom). At each $t_{\alpha}$ in the $* \to i$ time series, $\tau_{ij}$ is the time elapsed to the next $i \to j$ event, which is different from the inter-event time $\delta t_{ij}$ in the $i \to j$ time series. The red shaded area represents the recover time window $T_i$ after $t_{\alpha}$.}\label{figarrows}
\end{figure}

\begin{figure}
\includegraphics[width=0.45\textwidth]{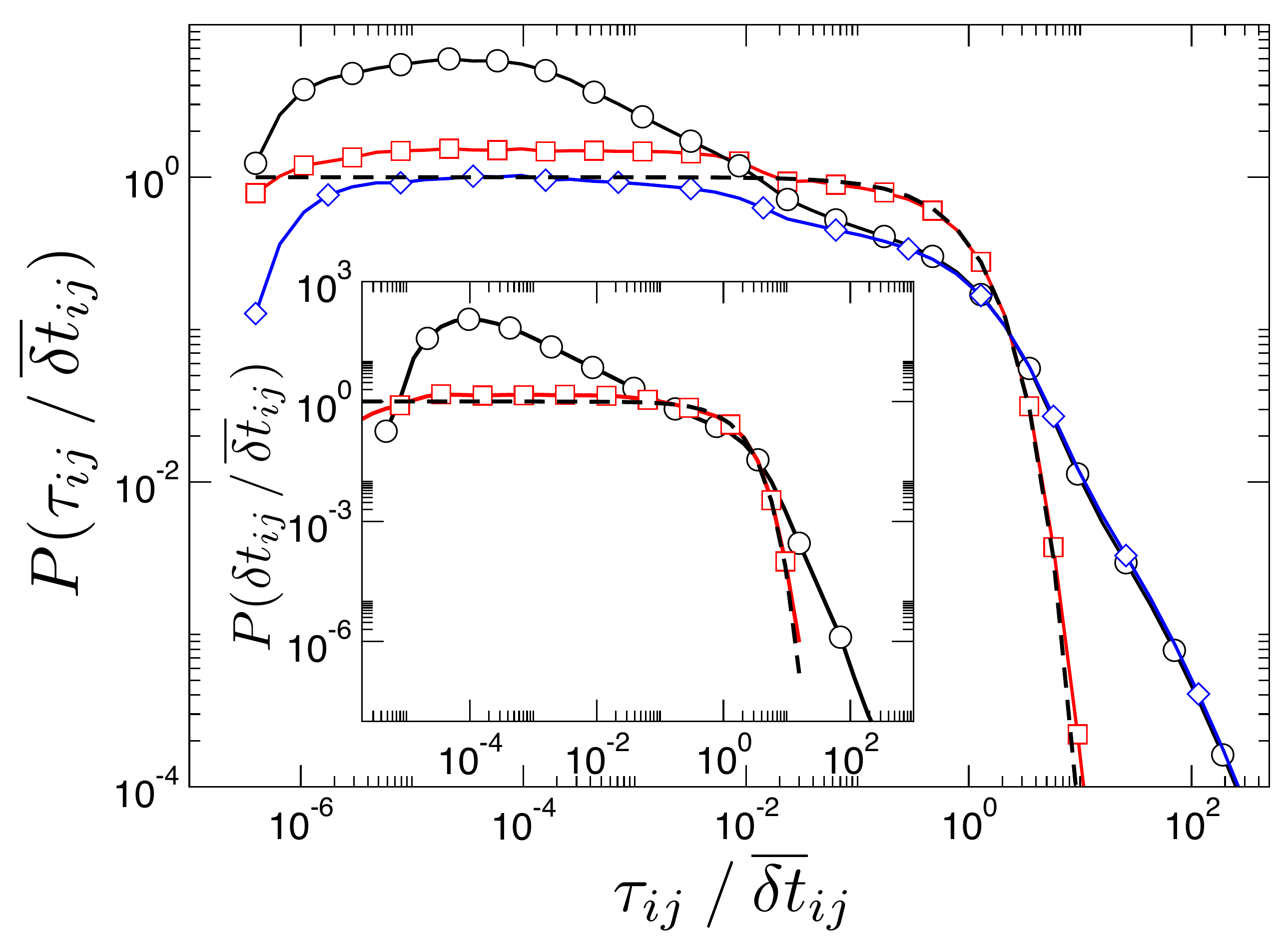}
\caption{(color online) Distribution of the relay time intervals $\tau_{ij}$ (main) and of the inter-event times $\delta t_{ij}$ (inset) in the $i \to j$ tie rescaled by $\overline{\delta t}_{ij}$. The black circles correspond to the  real data, while the red squares is the overall-shuffled result. Blue diamonds correspond to the case in which only the $* \to i$ sequence is randomized. Only ties with $w_{ij} \geq 10$ are considered. In both graphs the dashed line correspond to the $e^{-x}$ function.}\label{figpdf}
\end{figure}

Firstly, we investigate the communication temporal patterns that might affect information diffusion. 
Spreading from user $i$ to user $j$ ($i \to j$) happens at the {\em relay time} intervals $\tau_{ij}$ (also called inter-contact time \cite{isella}), i.e. the time interval it takes to $i$ to pass on to $j$ an information he/she got from any another person $* \to i$ (where $j\neq *$, see Fig.\ \ref{figarrows}). Information spreading is thus determined by the interplay between $\tau_{ij}$ and the intrinsic timescale of the infection process. Note that $\tau_{ij}$ depends on the correlated and causal way in which group conversations happen, since it depends on the inter-event intervals $\delta t_{ij}$ in the $i \to j$ communication but {\em also} on the possible temporal correlation with the $* \to j$ events \cite{newman}.
Ignoring this correlation, it is possible to approximate the probability distribution function (pdf) for $\tau_{ij}$ by the waiting-time density for $\delta t_{ij}$ \cite{vazquez,karsai} 
\begin{equation}\label{karsaipdf}
P(\tau_{ij}) = \frac{1}{\overline{\delta t}_{ij}} \int_{\tau_{ij}}^\infty P(\delta t_{ij}) d \delta t_{ij},
\end{equation}
where $\overline{\delta t}_{ij}$ is the average inter-event time. In this approximation, the dynamics of the transmission process only depends on the dyadic $i\to j$ sequence of communication events and in particular, the possible heavy-tail properties of $P(\delta t_{ij})$ are directly inherited by $P(\tau_{ij})$. Fig.\ \ref{figpdf} shows our (rescaled) results for $P(\delta t_{ij})$ and $P(\tau_{ij})$. For comparison, we also show the results obtained when {\em i)} the time-stamps of the $*\to i$ events are randomly selected from the complete CDR, thus destroying any possible temporal correlation with $i \to j$ and effectively mimicking Eq.\ (\ref{karsaipdf}) and {\em ii)} when the whole CDR time-stamps are shuffled thus destroying both tie temporal patterns and correlation between ties. Both shufflings preserve the tie intensity $w_{ij}$ \cite{onnela}, i.e.\ the number of calls and their duration and also the circadian rhythms of human communication \cite{karsai}. 
The result for $P(\delta t_{ij})$ shows that small and large inter-event times are more probable for the real series than for the shuffled ones, where the pdf is almost exponential as in a Poissonian process, apart from a small deviation due to the circadian rhythms. This bursty pattern of activity has been found in numerous examples of human behavior \cite{vazquez} and seems to be universal in the way a single individual schedules tasks. Here we see that it also happens at the level of two individuals interaction confirming recent results in mobile \cite{karsai} and online communities \cite{rybski} dynamics. The pdf for $\tau_{ij}$ is also heavy-tailed but displays a larger number of short $\tau_{ij}$ compared to the shuffled one. The abundance of short $\tau_{ij}$ suggests that receiving an information ($*\to i$) triggers communication with other people ($i \to j$), a manifestation of group conversations \cite{qiankun,eckmann,wu}. While the fat-tail of $P(\tau_{ij})$ is accurately described by Eq.\ (\ref{karsaipdf}), i.e.\ large transmission intervals $\tau_{ij}$ are mostly due to large inter-event communication times in the $i\to j$ tie,  the behavior of $P(\tau_{ij})$ is not only due to the bursty patterns of $\delta t_{ij}$, but also to the temporal correlation between the $i \to j$ and the $*\to i$ events. In fact, if the correlation between the $i\to j$ and the $* \to i$ series is destroyed, the probability of short-time intervals decreases and approaches the Poissonian case  (Fig.\ \ref{figpdf}).  
In summary, relay times depend on two main properties of human communication that compete to one another. While the bursty nature of human activity yields to large transmission times hindering any possible infection, group conversations translate into an unexpected abundance of short relay times, favoring the probability of propagation.

To investigate the effect of these two conflicting properties of human communication on information spreading, we simulate the epidemic Susceptible-Infectious-Recovered (SIR) model in our social network considering the real time sequence of communication events \cite{cebrian,karsai} and compare them to the shuffled data. We start the model by infecting a node at a random instant and considering all other nodes as susceptible. In each call an infected node can infect a susceptible node with probability $\lambda$. Due to the synchronous nature of the phone communication, this happens regardless of who initiates the call. However, since the same results are obtained by considering directionality in the calls, for computational reasons we consider the latter case. Nodes remain infected during a time $T_i$ until they decay into the recovered state. For the sake of simplicity we simulate the simplest model in which the recovering time $T_i$ is deterministic and homogeneous $T_i = T$ and set $T = 2$ days, although different and/or stochastic $T_i$ can be studied within the same model. The spreading dynamics generates a viral cascade that grows until there are no more nodes in the infected state. We repeat the spreading process for $3 \times 10^4$ randomly chosen seeds. Note that our model includes the SI model simulations in \cite{karsai} where $\lambda = 1$ and $T = T_0$, with $T_0$ being the total duration of the dataset. 

\begin{figure}
\includegraphics[width=0.48\textwidth]{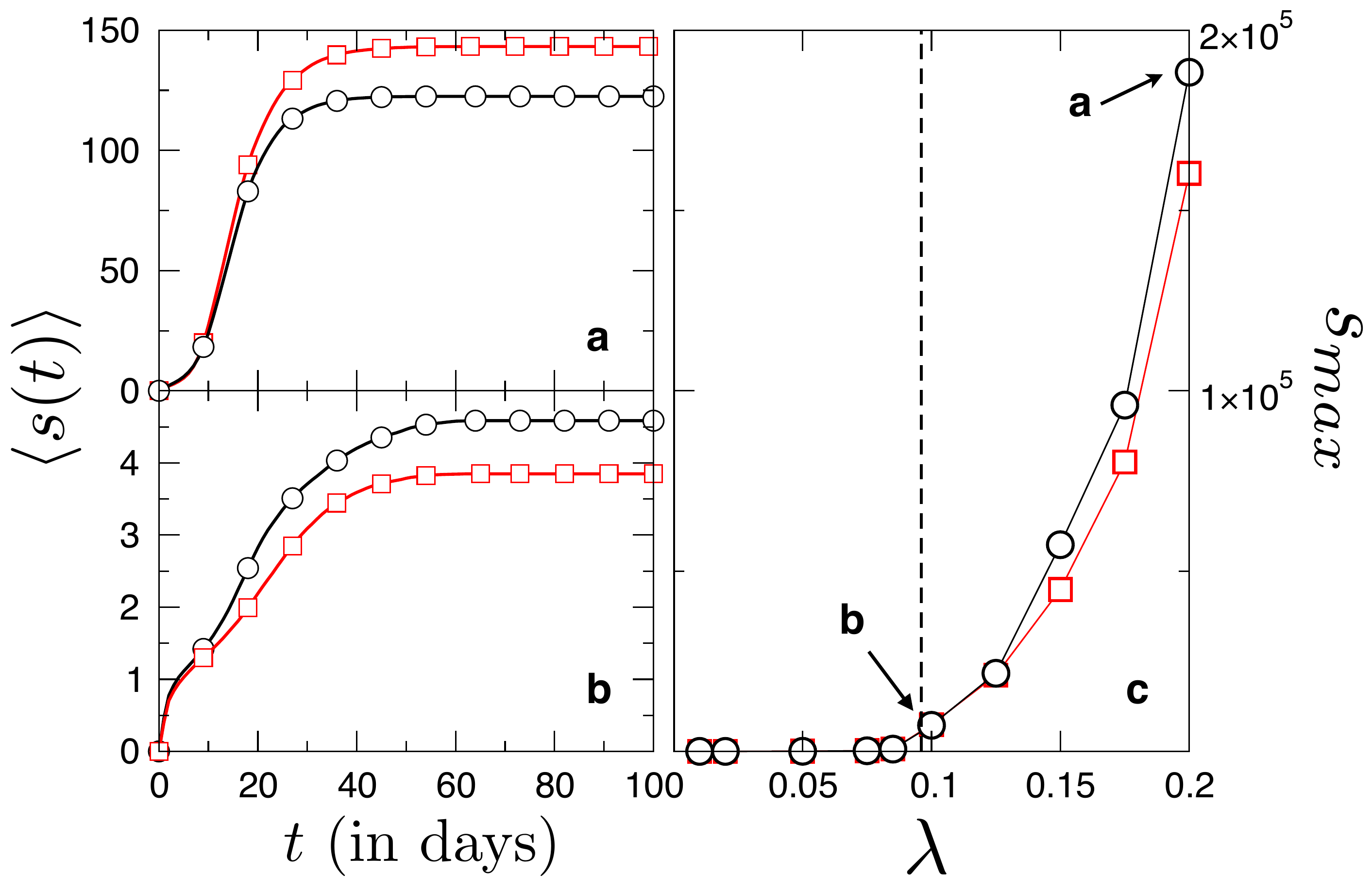}
\caption{(color online) Average size dynamics for a large (a) and a small (b) value of $\lambda$ (left) and maximum size (right) of the infection outbreaks (over $10^4 $ realizations) for the real data (black lines) and shuffled data (red lines) for $T = 2$ days. The dashed line shows the critical point estimation of the percolation transition given by $R_1[\lambda,T] = 1$ with $R_1$ calculated using Eq.\ (\ref{eqr1}).}\label{figperc}
\end{figure}

By looking at the size of the largest cascade $s_{max}$ (over all realizations) at each value of $\lambda$, we first ensure the existence of a percolation transition \cite{newmansiam} (see Fig.\ \ref{figperc}), confirmed by a change in the behavior of $s_{max}$ from small to large cascades at a given value of $\lambda$ (tipping point).
The same behavior is observed for the shuffled-time data where the transition seems to happen almost at the same value of $\lambda$, although an accurate analysis of the percolation point if beyond of the scope of this letter.
On the contrary, there is a significant difference in the behavior of the asymptotic average size $s_\infty$ between the real and the shuffled-time data for different regimes of $\lambda$: when $\lambda$ is small, $s_\infty$ is bigger for the real data than for the shuffled one, while the opposite behavior is observed for large $\lambda$. This difference, that can be very large for moderate values of $\lambda$, shows the impact of the real time dynamics of communication in the reach of information in society. Specifically, if information propagates easily (large $\lambda$), the average reach in social networks is narrower than the one expected when a Poissonian dynamics is considered. In this sense, temporal patterns make social networks bigger than expected at large scales. 
However, in most real situations $\lambda$ is very small \cite{esteban} and in this case the observed behavior is the opposite: despite the low propagation, information cascades are larger in real data than in the Poisson case, which suggests that information spreading is more efficient at small (local) scales.

\begin{figure}
\includegraphics[width=0.48\textwidth]{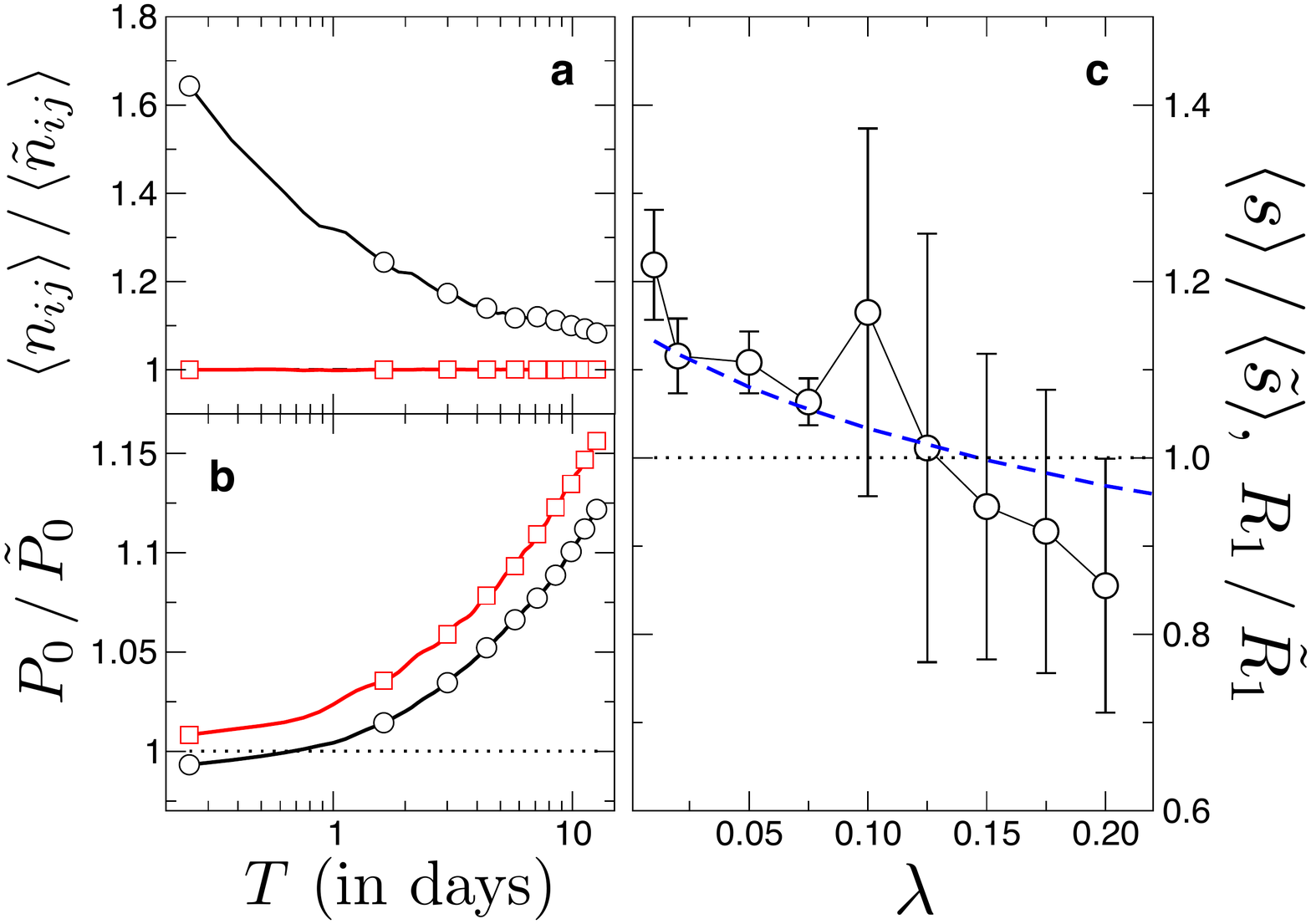}
\caption{Ratio of the number of events ({\bf a}) and probability of no events ({\bf b}) as a function of the recovery time $T$ for the real (black circles) and shuffled $* \to i$ (red squares) data with respect to the overall-shuffled data. Right panel ({\bf c}) shows the ratio of the average size of the outbreaks (black circles) and of $R_1$ calculated using Eq. \ref{eqr1} (dashed blue line).}\label{figrelsize}
\end{figure}

To understand this behavior, we follow the approach of \cite{newman} mapping the dynamical SIR model to a static edge percolation model where each tie is described by the transmissibility ${\cal T}_{ij}$, that represents the probability that the information is transmitted from $i$ to $j$ and is a function of $\lambda$ and $T$.
If user $i$ becomes infected at time $t_\alpha$ and the number of communication events $i\to j$ in the interval $[t_\alpha,t_\alpha+T]$ is $n_{ij}(t_\alpha)$, then the transmissibility in that interval is (see Fig.\ref{figarrows}) ${\cal T}_{ij} = 1 - (1-\lambda)^{n_{ij}(t_\alpha)}$.
User $i$ may become infected at any $* \to i$ communication event. 
Assuming these events independent and equally probable, we can average ${\cal T}_{ij}$ over all the $t_\alpha$ events to get
\begin{equation}\label{eqT}
{\cal T}_{ij}[\lambda,T] = \langle 1-(1-\lambda)^{n_{ij}(t_\alpha)}\rangle_{\alpha}.
\end{equation} 
If the number of  $* \to i$ events is large enough we could use a probabilistic description of Eq.(\ref{eqT}) in terms of the probability $P(n_{ij}=n;T)$ that the number of communication events between $i$ and $j$ in a given time interval $T$ is $n$. Thus
\begin{equation}\label{tp}
{\cal T}_{ij}[\lambda,T] = \sum_{n=0}^\infty P(n_{ij}=n;T) [1-(1-\lambda)^n],
\end{equation}
which in principle can be non symmetric (${\cal T}_{ij} \neq {\cal T}_{ji}$).
This quantity represents the real probability of infection from $i$ to $j$ and defines  the {\em dynamical strength} of the tie. Note that ${\cal T}_{ij}$ depends on the series of communication events between $i$ and $j$, but also on the time series of calls received by $i$. In \cite{newman} Newman studied the case in which both time series are given by independent Poisson processes in the whole observation interval $[0,T_0]$. Thus, $P(n_{ij}=n;T)$ is the Poisson distribution with rate $\rho_{ij} = w_{ij} T / T_0$, where $w_{ij}$ is total number of calls from $i$ to $j$ in $[0,T_0]$, thus
\begin{equation}
\tilde {\cal T}_{ij}[\lambda,T] = 1 - e^{-\lambda \rho} = 1 - e^{-\lambda w_{ij}T /T_0},
\end{equation}
which shows the one-to-one relationship between the intensity $w_{ij}$ and the transmissibility ${\cal T}_{ij}$ in the Poissonian case: the more intense the communication is, the larger the probability of infection. 
However, as we have seen in Fig.\ \ref{figpdf}, the real $i \to j$ and $*\to i$ series are far from being independent and Poissonian and in order to investigate the effect of real patterns of communication on the transmissibility we approximate Eq.\ (\ref{eqT}). For small values of $\lambda$ we have $1-(1-\lambda)^n \simeq \lambda n$, while when $\lambda \simeq 1$ we get that $1-(1-\lambda)^n \simeq 1$ for $n > 0$. Thus, the transmissibility for the two regimes is given by:
\begin{equation}
{\cal T}_{ij}[\lambda,T] = \left \{
\begin{array}{ll}
\lambda \langle n_{ij}\rangle_{t_\alpha} & \mathrm{when}\ \lambda \ll 1\\
1 - P_{ij}^0 & \mathrm{when}\ \lambda \simeq 1
\end{array}\right.
\end{equation}
where $P_{ij}^0 = P(n_{ij}=0;T)$. Specifically, $P_{ij}^0$ can be estimated directly from Eq.\ (\ref{karsaipdf}) for each link $P_{ij}^0 = \int_T^\infty P(\tau_{ij}) d\tau_{ij}$, since it measures the probability to find a relay time bigger than $T$.  Fig.\ \ref{figrelsize} shows the comparison of $n_{ij}$ and $P_{ij}^0$ (averaged over all links) for different values of $T$ for the real and shuffled data (denoted by tilde). On one side, due to the correlation between the $*\to i$ and $i \to j$ time series, the number of events in a tie following an incoming call is always larger for the real data than for the shuffled one. This is the reason why, for small $\lambda$, the average transmissibility (and thus the size of the epidemic cascades) is always higher in real communication patterns \cite{qiankun}.
On the contrary, the bursty nature of the $i\to j$ communication makes the tail for the real $P(\tau_{ij})$ heavier than the exponential distribution found in the shuffled data. Thus if $T$ is large enough, $P_{ij}^0$ is larger in the real than in the shuffled data and this is why we observe smaller cascades in that region.
Note however that this does not apply for very small values of $T$ ($T \lesssim 1$ days), where the causality between $* \to i$ and the $i\to j$ time series can make $P_0$ even smaller in the real case.

To give a more quantitative analysis of the observed behavior we investigate the percolation process in a social network in which links have transmissibility ${\cal T}_{ij}$. The important quantity is the secondary reproductive number $R_{1}$, that is the average number of secondary infections produced by an infectious individual. $R_{1}$ gives information about percolation transition in the SIR process (which happens at $R_1 = 1$ \cite{newman}), but also about the speed of diffusion (which is proportional to $R_1$ \cite{barthalemy}) and of the size of the cascades (which is a growing function of $R_1$ \cite{newman}). Assuming that the ${\cal T}_{ij}$ are given and that the social network is random in any other respect, $R_{1}$ can be approximated as
\begin{equation}\label{eqr1}
R_{1}[\lambda,T] = \frac{\langle (\sum_j {\cal T}_{ij})^2\rangle_i - \langle \sum_j {\cal T}_{ij}^2 \rangle_i}{\langle \sum_j {\cal T}_{ij}\rangle_i}.
\end{equation}
Note that in the homogeneous case in which ${\cal T}_{ij} = {\cal T}$ we recover the common result in random networks $R_1 = {\cal T}(\langle k_i^2 \rangle/\langle k_i \rangle -1)$ \cite{newman}. Figs.\ \ref{figperc} and \ref{figrelsize} show the accuracy of the approximations used to get Eq.(\ref{eqr1}) to predict the tipping point in the SIR process and the change in the average size of the cascades in the two regimes.
This suggests that the dynamical strength of the ties ${\cal T}_{ij}$, defined in Eq.\ (\ref{eqT}), can be effectively used to model real strength of human interactions in social networks. 

In conclusion we have seen that both the bursty nature of human communications and the existence of group conversations are the two main dynamical ingredients to understand the spreading of information in social networks. These two effects compete in the spreading, favoring and hindering information reach when compared with the homogeneous case. Our results indicate the necessity to incorporate temporal patterns of communication in the description and modeling of human interaction. Actually, we have proven an effective way to map the dynamics of human interactions onto a static representation of the social network through the concept of {\em dynamical strength} of ties. We believe its success in explaining information diffusion would encourage the use of this {\em dynamical strength} in other areas of network research which is based on information spreading like the determination of influence/centrality \cite{newman2}, community finding \cite{community}, viral marketing \cite{esteban,cebrian}, etc.

We thank J.\ L.\ Iribarren and R. Cuerno for discussions and Telef\'onica for the access to anonymized data. G.\ M.\ and E.\ M.\ acknowledge support from MEC (Spain) through grants i-MATH and MOSAICO.

\end{document}